\documentclass[letterpaper,aps,prl,twocolumn,showpacs,preprintnumbers]{revtex4}
\usepackage{graphicx}
\begin{document}

\newcommand{\mysw}[1]{{\scriptscriptstyle #1}}
\newcommand{\mybm}[1]{\mbox{\boldmath$#1$}}
\preprint{\sf Version 2 (\today)}
\title{Direct Measurement of Spatial Distortions of Charge Density Waves in K$_{0.3}$MoO$_3$}
\author{Chao-Hung Du$^{1,2}$, Mau-Tsu Tang$^{2}$, Yen-Ru Lee$^{3}$, Yuri P. Stetsko$^{2,3}$, Chung-Yu
Lo$^{3}$, Jey-Jau Lee$^{2}$, Hsiu-Hau Lin$^{3,4}$, and Shih-Lin
Chang$^{2,3}$}

\affiliation{$^{1}$Department of Physics, Tamkang University, Tamsui 25137, Taiwan\\
$^2$National Synchrotron Radiation Research Center, Hsinchu 300, Taiwan\\
$^3$Department of Physics, National Tsing Hua University, Hsinchu 300, Taiwan\\
$^4$Physics Division, National Center for Theoretical Sciences,
Hsinchu 300, Taiwan }
\date{\today}

\begin{abstract}
Using X-ray scattering and the technique of multiple diffreactions, we revisit the dynamical transition of charge density waves (CDWs) in K$_{0.3}$MoO$_{3}$ under applied voltages. In addition to the usual transport and half width (of Bragg peaks) measurements, we also measure the triplet phase by three-wave diffraction, which provides, for the first time, the {\em direct} evidence for the spatial distortions of CDWs. This novel and sensitive technique developed here can be applied to general periodic media, including stripes in high temperature superconductors, and provide a new perspective into interesting phenomena in these materials.
\end{abstract}

 \maketitle

The homogeneous phases in low dimensional materials, such as
K$_{0.3}$MoO$_{3}$, NbSe$_{3}$ and TMTSF molecules, undergo a
phase transition to charge density waves (CDWs) at low
temperatures. The instability toward spontaneous formation of
charge-density modulations is driven by electron-phonon
interactions, or sometimes electron-electron ones
\cite{Overhauser78}. Among many other interesting aspects of CDW,
transport property in the presence of finite driving electric
field has attracted lots of attentions from both experimental and
theoretical sides. It is generally believed that the current is
suppressed at small biased voltage, where the CDW is pinned by
impurity potential. Above some threshold voltage, the sliding
motion along the applied electric field starts and the current
increases significantly. It has been known that this dynamic
behaviour involves the phase slippage of the density waves. In
fact, similar phenomena occurs in many other systems, such as
moving vortex lattice \cite{Yaron95}, Wigner crystal
\cite{Andrei88}, charge/spin stripes in CMR, high-T$_{C}$
superconductors
 \cite{Yaron95,Xiao01} and La$_{2-{\it x}}$Sr$_{\it x}$NiO$_{4}$
\cite{Littlewood95,Du00}, colloids \cite{Murray90}, magnetic
bubbles \cite{Seshadri92} and so on.

We revisit this well-studied transition in this Letter, attempting
to measure the distortions of CDW in pinned and sliding phases
directly. In previous studies, the sliding transition is spotted
by sudden jump in $I-V$ characteristics at low temperatures
\cite{Ogawa01} and/or the half-width change of the Bragg peaks
\cite{Danncau02}. Theoretical investigations
\cite{Giamarchi95,Balents98} predicted that the pinning forces
become irrelevant when the system enters the sliding phase. This
indeed provides a natural explanation for the dynamical narrowing
of the half width above the threshold voltage. However, to
understand how the CDW adjusts to the pinning forces at different
driving voltages, a direct measurement for the spatial distortions
of CDW is desirable. In this Letter, in addition to the usual
transport and FWHM (full width at half maximum) measurement of the
CDW satellite reflections, we first establish the connection
between the lattice distortions and the triplet phase in X-ray
scattering and thus demonstrate how the spatial distortions of CDW
can be measured directly. While the sliding transition is already
well studied, the technique we developed here can be applied to
general periodic media driven by external sources and provides a
new perspective into many interesting strongly correlated systems.

\begin{figure}
\centering
\includegraphics[width=6cm]{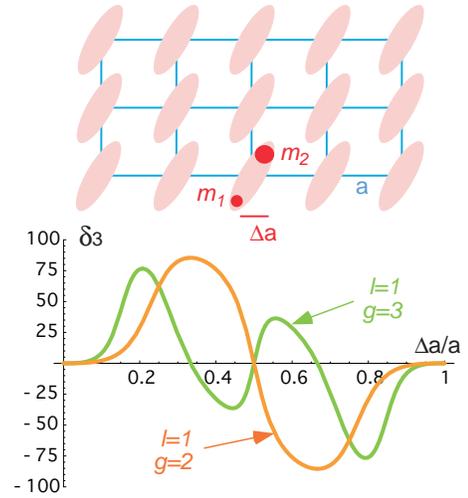}
\caption{\label{Fig1} (Color online) (a) (up) Two-dimensional
square lattice with distortion $\Delta a$, (b) (bottom) Triplet
phase $\delta_3$ versus the distortion $\Delta a/a$. Two specific
choices $l=1$, $g=2$ and $l=1$, $g=2$ are presented here. }
\end{figure}

Since the triplet phase $\delta_3$ plays an essential role in our
study, it is helpful to illustrate its physical meaning and
connections to lattice distortions first. The triplet phase $\delta_3$
is defined as the phase of the structure factor triplet $F_{G_2}
F_{G_3}/F_{G_1}$, where $G_i$ are reciprocal lattice vectors, satisfying $G_1=G_2+G_3$. \cite{Chang04,Colella74,Shen98}. Since they form a closed triangle in the reciprocal vector space, it is
straightforward to show that the triplet phase $\delta_3$ is
invariant under arbitrary choices of unit cells \cite{Giaco02}.
This invariance provides a hint for its connection to some
physical quantity, which turns out to be the internal distortion
of the unit cell. Experimentally, this triplet phase can be
determined by measuring the diffracted intensity profiles of a
three-wave $(O, G_1, G_2)$ diffraction involving the primary $G_1$,  the secondary $G_2$, and the coupling $G_3=G_1-G_2$ reflections, where $O$ stands for the direct reflection of the incident beam \cite{Chang04}.  
Let's consider a two-dimensional square
lattice with two ions in one unit cell, as shown in
Fig.~\ref{Fig1}. Without distortion $\delta a=0$, the lattice has
inversion symmetry which ensures all structure factors are real
(under appropriate choice of the unit cell) and thus gives
$\delta_3 =0$. Under the driven voltage, the CDW is distorted and
twists the underlying lattice as well. For simplicity, let's
assume that it can be described by a lattice twist $\delta a$
along direction of CDW ($x$-axis here). Furthermore, let's choose
the reciprocal lattice vectors to be along the direction of CDW,
i.e. $G_1 = (2g\pi/a, 0)$, $G_2 = (2l\pi/a,0)$ and $G_3 =
G_1-G_2$. The resultant triplet phase can be computed
straightforwardly,
\begin{eqnarray}
\tan \delta_3 = \frac{\Delta \sum_{i} \sin(G_i \delta a+\varphi_i)}{r + \sum_{i} \cos(G_i \delta a)},
\end{eqnarray}
where $\varphi_1=\pi$ and $\varphi_{2,3}=0$. The other parameters are $\Delta = (m_1 -m_2)/(m_1+m_2)$ and $r =
(m_1^3+m_2^3)/2m_1m_2(m_1+m_2)$. Two particular choices of $l$ and
$g$ are given in Fig.~\ref{Fig1} to demonstrate the connection
between the triplet phase $\delta_3$ and the lattice distortion
$\delta a$.

In realistic setup, the lattice distortion is rather small $\delta
a/a \ll 1$ at all applied voltages, thus the expression of the
triplet phase simplifies, $\delta_3 \sim (\delta a/a)^3$. Note
that the cubic dependence is {\em generic} due to the inversion symmetry
of the undistorted lattice and the close triangle formed by $G_i$. While
this result is derived from the simple model, it captures the
generic dependence of the triplet phase even for the more
complicated crystal K$_{0.3}$MoO$_{3}$ we studied here. Therefore,
the triplet phase provides a direct measurement of the spatial
distortions of CDWs at different driving voltages.

\begin{figure}
\centering
\includegraphics[height=6cm]{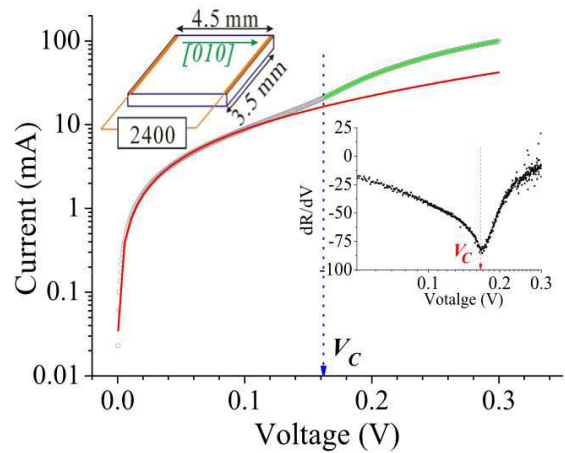}
\caption{\label{Fig2} (Color online) $I-V$ characteristic of
K$_0.3$MoO$_3$ in a two-probe transport setup at $T = 70$ K. The
red line is the fit to the predicted thermal creeps when the CDW
is pinned. The left inset shows the dimension of the sample and
the experimental setup. The right inset for $(dR/dV)$ shows a
transition point at $V_c$= 0.165 V.}
\end{figure}

Now we turn to the experimental details and the observed results
of transport and X-ray measurements. A single crystal
K$_{0.3}$MoO$_{3}$ of good quality was prepared for the transport
measurement and X-ray scattering. The crystal structure belongs to
the monoclinic with the space group {\it C2/m}. The lattice
parameters of K$_{0.3}$MoO$_{3}$ are $a =18.162$ \AA, $b = 7.554$
\AA, $c = 9.816$ \AA, and $\beta$ = 117.393$^{\circ}$
\cite{Schutte93}.  The sample, with transition temperature around 180 K, was characterized with a mosaic width of ~0.005$^{\circ}$ and pre-aligned using an X-ray rotating anode source so that the scattering plane coincided with $ a^* \times c^*$ plane.
The in-situ measurements were carried out on the Taiwan beamline
BL12B2 of SPring-8 synchrotron facility. The incident x-ray
wavelength was selected to be 1\AA. Two gold stripes spaced about
3 mm were evaporated onto the sample surface as shown in the inset
of Fig.~\ref{Fig2}. The sample was then glued on the cold head of
a cryostat mounted on a 6-circle diffractometer. To drive current
through the sample, the voltage was applied along $b^*$ axis
([010] direction).  A Keithely 2400 source meter was used to
generate the driving voltage, and the $I-V$ curve was measured by
the two-probe setup. An upper limit of the current was set to 300
mA in order to protect the sample and meter.

Fig.~\ref{Fig2} shows the non-linear conductivity of the sample at
T = 70 K, indicating the dynamical transition from the pinned CDWs
to sliding motions. While the nonlinearity is not as robust as at
low temperatures, the current below the critical voltage $V_c
\approx 0.165$ V can be fitted remarkably well with the prediction
from thermal creep
\cite{Ogawa01,Chauve00,Marchetti03,Fisher83,Chen96},
\begin{eqnarray}
I(V) = G_0 (V-V_0) \exp \left(\alpha \frac{V}{T} \right).
\end{eqnarray}
The parameters $G_0$, $V_0$ and $\alpha$ are constants at all
applied voltages. To make the critical transition more
transparent, one can plot $dR/dV$ (as shown in the inset of
Fig.~\ref{Fig2}), which show clear singularity near the critical
voltage $V_c$. The nice fit with the thermal creep behavior
indicates that our two-probe measurement does not suffer poor
contacts or serious current inhomogeneity in the sample. One may
notice that there is no switching phenomena in our measured $I-V$
curve due to thermal fluctuations at $T=70$ K
\cite{Vinokur97,Adelman93}. 

\begin{figure}
\centering
\includegraphics[height=8cm, width=7cm]{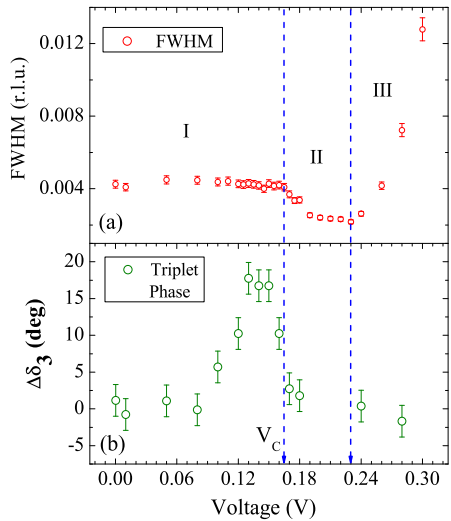}
\caption{\label{Fig3} (Color online) (a) Evolution of the half
width of the Bragg peak versus the applied voltage. According to
the changes of the half width, one can classify CDW into three
phases: (I) the creeping CDW state, (II) the moving solid, and
(III) the moving liquid. (b) The triplet phase change
$\Delta\delta_3$ at different voltages. Note that, in the sliding
phase, $\Delta\delta_3 = 0$ is a direct evidence that the pinning
forces become irrelevant.}
\end{figure}

In addition to the transport measurement, the evolution of CDW satellite reflections as a function of applied fields was also probed using X-ray scattering. The width of the CDW peak and triplet phase at different driving
voltages are summarized in Fig.~\ref{Fig3}. In this Letter, we
focus on the particular satellite reflection, located at $G_{1}$ =
(13 $q$ -6.5) with $q \approx 0.748$. Scans were performed along
the longitudinal direction of [2 0 -1] and the data were
convoluted with resolution function obtained from nearby Bragg
peak (12 0 -6). The FWHM of the primary reflection $G_1$ in Fig.~\ref{Fig3}(a) remains more or less unchanged below
the critical voltage $V_c \approx 0.165$ V, determined from the
transport measurement. Above the critical voltage, where the CDW
enters the sliding phase, the FWHM decreases\cite{Koshelev94,Reichhardt03} as predicted by
previous theoretical investigations\cite{Koshelev94}. This interesting dynamical narrowing of half width is a strong indication that the pinning
forces due to random potentials become irrelevant (or less
efficient) when the CDW starts to slide\cite{Chen96,Lemay99,Danncau02}.
In fact, similar motion-induced ordering behavior by external driving force has been reported in NbSe$_{3} as well$ \cite{Danncau02}. When the voltage goes beyond 0.22 V, other effects from non-equilibrium dynamics and amplitude fluctuations become important and the Bragg peaks disappear. This is clearly evidenced by the sharp increase of FWHM in Fig~\ref{Fig3}(a) and diminishing amplitude of the corresponding Bragg peak (not shown here).

Now we come to the central quantity we studied in this Letter -- the triplet
phase  $\delta_{3}$ of the 3-wave multiple diffraction at different
biased voltages. To set up a 3-wave $(O,G_1,G_2)$ multiple
diffraction experiment, where $O$ stands for the direct incident reflection,
the crystal is first aligned for a primary reflection $G_1$. It is then rotated around the reciprocal lattice vector $G_1$ with an azimuthal angle $\psi$ to bring in the secondary reflection $G_2$ which also satisfies Bragg's law.
Namely, both $G_1$ and $G_2$ reflections take place simultaneously.  The
interaction of the multiply diffracted waves modifies the intensity of the primary reflection. Intensity variation showing asymmetric distribution versus  $\psi$, giving the information about the triplet phase $\delta_3$ of the three-wave structure factor $F_{G_2}F_{G_3}/F_{G_1}$\cite{Chang04,Colella74,Shen98}. Previously \cite{Du04}, we showed how the triplet phase $\delta_{3}$ can be
probed using multiple diffraction. Here we further demonstrate
that measuring the change of the triplet phase $\Delta\delta_{3}$ allows us to make direct observation of the internal deformations of the CDWs/lattice at different driving voltages.

\begin{figure}
\centering
\includegraphics[height=6cm]{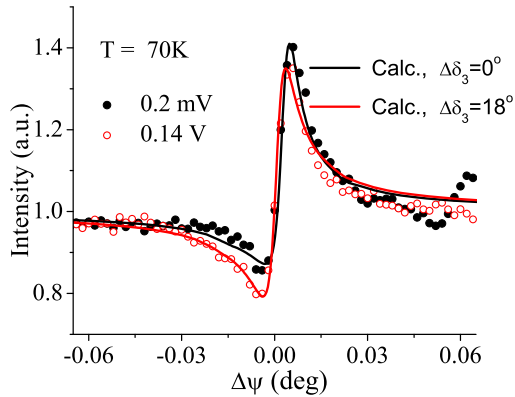}
\caption{\label{Fig4} (Color online) Triplet phase change
$\Delta\delta_3$ extrapolated from curve fitting of the 3-wave
diffraction profiles for $V$ = 0.2 mV and 0.14 V: Since the
primary (13 $q$ -6.5) and the coupling reflection (9 $q+8$ 10.5)
are the fractional reflections, their structure factor amplitudes
are much smaller than that of the secondary (4 -8 4) reflection.
Also the amplitudes of (13 $q$ -6.5) are nearly the same for $V <
0.18$ V. Under this condition, the modification of profile
asymmetry is dominated by the phase, rather than the amplitude of
the structure-factor triplet. The analysis is based on the dynamic
theory for multiple diffraction, giving   $\Delta\delta_{3}$ =
0$^{\circ}$ and  18$^{\circ}$ at $V$ = 0.2 mV and 0.14 V,
respectively.}
\end{figure}

The origin of azimuthal angle ($\psi$=0) was determined to be the
direction where [1 0 0] lay on the scattering plane. This can be
verified by finding a mirror position in the multiple diffraction
pattern of the primary reflection (6 0 -3). Through the azimuthal
scan around the primary reflection $G_1$ = (13 $q$ -6.5) at $T$ = 70
K, we obtained three-wave diffraction pattern containing lots
of multiple diffraction peaks\cite{Lo04}. In this Letter, we concentrated
only on the particular three-wave diffraction, $G_1$ = (13 $q$
-6.5) and $G_2$ = (4 -8 4) at $\psi$=108.53$^{\circ}$. Note that the coupling reflection is $G_3 = G_1-G_2$ = (9 $q$+8 -10.5). The profile asymmetry of the diffraction intensity of $G_1$ versus $\psi$ at V= 0 is typical because the triplet phase $\delta_3=0$ due to the centrosymmetry of the undistorted lattice. Upon application of driving voltage, the change of the triplet phase $\Delta \delta_3$ is analyzed based on the dynamical theory for multiple
diffractions\cite{Chang04,Colella74,Shen98}. In Fig.\ref{Fig4}, the
profiles with typical asymmetry ($\Delta \delta_3 =0$) and the distorted one ($\Delta \delta_3 = 18^{\circ}$) are displayed. Note that the change in the peak profile for $\Delta \delta_3 = 18^{\circ}$ indicates that the original centrosymmetry is broken due to lattice deformation caused by ion displacements under the influence of external forces. The measured $\Delta \delta_3$ in Fig.~\ref{Fig3}(a) is very similar to the simplified calculation of triplet phase in Fig.~\ref{Fig1} when the deformation is small $\Delta a/a \ll 1$. (Note that, in the present case, $\Delta \delta_3 =\delta_{3}$ because the original value of $\delta_{3}$ is zero for the undistorted lattice.) The
Darwin width of a Bragg reflection was also monitored in order to
make sure the crystal was not destroyed by the applied voltage. As
shown in Fig.~\ref{Fig3}(b), the change of the triplet phase $\Delta\delta_{3}$ hits its maximum at $V$ = 0.12 $\sim$ 0.14 V. This clearly shows that the internal distortions of CDWs reach saturation just before the sliding motion.
Loosely speaking (ignoring spatial inhomogeneity), at low bias voltages, the free energy is minimized by the small cost of elastic energy due to CDW distortions. After threshold voltage, it is energetically favorable to slide
(costing kinetic energy) rather than holding up the large elastic
energy.  The estimated $\Delta\delta_{3}$ from curve fitting at
0.1, 0.12 0.13, 0.14, 0.15, and 0.16 V are about 6$^{\circ}$,
10$^{\circ}$, 18$^{\circ}$, 17$^{\circ}$, 17$^{\circ}$, and
10$^{\circ}$ respectively, and then back to 0$^{\circ}$ for $V>V_c$. As far as we know, this is the first {\em direct} observation of the spatial distortions of CDWs from the pinned to sliding states.\cite{Gruner94,Brazovskii00}

In summary, we demonstrate the simultaneous measurements of transport, half-width of the Bragg peak and the triplet phase of CDWs in K$_0.3$MoO$_3$ at different driving voltages. The combination of different measurements provides evidence for the origin of the nonlinearity and dynamical phase transition in periodic media In particular, the phase measurement using three-beam diffraction is demonstrated for the first time as a novel and sensitive method to probe dynamical phenomena in nonlinear systems. While it is already exciting to observe how the internal deformations respond through the dynamical transition of CDW, it also opens up many interesting issues requiring further studies. For instance, the powerful technique developed here can be use to study the crossovers/transitions between different types of dynamical phases and a global phase diagram at different temperatures and voltages can be mapped out without ambiguity. This technique can also been applied to general periodic media, such as stripes/CDWs in high-$T_{c}$ related perovskites, and possibly deepen our understanding of dynamical motions in these materials.

The authors acknowledge the Ministry of Education, National
Science Council in Taiwan and the National Synchrotron Radiation Research
Center for financial supports, through the grants No. 90-FA04-AA,
NSC 91-2112-M-213-016, NSC 92-2112-M-032-013 and NSC93-2112-M-007-005. The beam time arrangements by the NSRRC and SPring-8 are also gratefully appreciated.

\end{document}